\documentclass[
     ,final            
  ]
  {aipproc}
  
\usepackage{floatflt} 

\layoutstyle{8x11single}

\begin{document}

\title{ Determination of PDFs and {\boldmath $\alpha_{S}(M_{Z})$} from Inclusive and Jet Measurements in DIS @ HERA }



\author{ G\"unter Grindhammer
 $\,$ (for the H1 and ZEUS collaborations) }{
  address={ Max-Planck-Institute f\"ur Physik, F\"ohringer Ring 6, 80805 Munich, Germany$\;$(mail:guenterg@desy.de) }
}



\begin{abstract}

A simultaneous NLO QCD fit of the parton density functions (PDFs) of the proton and of the strong coupling $\alpha_{S}(M_{Z})$ is presented. The analysis is based on combined H1 and ZEUS inclusive DIS cross section measurements together with inclusive jet cross section measurements provided by both H1 and ZEUS collaborations. The inclusion of the jet data significantly reduces the correlation between the gluon PDF and the strong coupling, improving the precision of the gluon PDF and providing an accurate determination of $\alpha_{S}(M_{Z})$ using DIS data from HERA only.
\end{abstract}

\maketitle


\section{Introduction}
\label{sect:intro}
The Rutherford centennial is celebrating Rutherford's discovery of the atomic nucleus in 1911 by scattering alpha-particles on a gold foil. Since then scattering experiments have come a long way, improving the resolution from about $6$ times the proton radius ($R_{p} \approx 1$~fm) in 1911 to about $1/1000 \cdot R_{p}$ as achieved by the HERA experiments \cite{h1-contact-interactions} at the so far only $ep$ collider. Here, I want to describe the current achievements in the extraction of the parton density functions (PDFs) of the proton and of the strong QCD coupling from inclusive and jet measurements in DIS at HERA.

Inclusive neutral current (NC) and charged current (CC) measurements, in leading order (LO) represented by the quark-parton model, are sensitive to the valence and sea quarks of the proton. Only at next-to-leading order (NLO) and beyond do they become sensitive to the gluon density (scaling violations) and the strong coupling, which are however strongly correlated. Inclusive jets in DIS, when measured for example in the Breit frame, are sensitive to the gluon density and the strong coupling already in LO, via the boson-gluon fusion process. This process dominates for the negative photon momentum transfer squared, or photon virtuality, $Q^{2}$, less than about $1000$~GeV$^{2}$. Above that value the QCD-Compton process becomes dominant, providing sensitivity to the strong coupling without being correlated to the gluon. Therefore, one can expect that the addition of jet measurements to the hitherto mainly used inclusive measurements alone will allow a reduction of the correlation between the gluon density and the strong coupling, thereby making it possible to extract simultaneously both PDFs and the strong coupling from DIS data.

\section{HERAPDFs}
The idea behind the HERAPDFs is to use only HERA NC and CC data in the fits. Combined inclusive cross sections from the H1 and ZEUS experiments are used, which have increased precision due to statistics but also due to intercalibration effects made possible by the two complementary detectors. Thus, precise combined data sets with total uncertainties between $1-2\%$ (for NC) over most of the phase space are used. The systematic correlated and uncorrelated errors are well controlled, such that the standard $\Delta\chi^{2}=1$ uncertainty criterion can be used. Since HERA data only include $e^{\pm}p$ data, there are no uncertainties due to for example deuterium or heavy target corrections.

Two parametrizations have been studied: a $10$ parameter fit which was used for the published HERAPDF1.0 \cite{herapdf10} and for the preliminary HERAPDF1.5 \cite{herapdf15}, and a $14$ parameter fit, for the preliminary HERAPDF1.5f and HERAPDF1.6 \cite{h1-zeus-pdf-alphas-fits}. The latter parametrization provides more flexibility, in particular for the gluon density at low $x$, which is preferred because of the additional jet data. At the $Q_{0}^{2}$ starting scale the PDFs are parametrized as follows:
\begin{eqnarray*}
xg(x) & = & A_{g}x^{B_{g}} \cdot (1-x)^{C_{g}} - A_{g}^{\prime}x^{B_{g}^{\prime}} \cdot (1-x)^{C_{g}^{\prime}} \\
xu_{v}(x) & = & A_{u_{v}} x^{B_{u_{v}}} \cdot (1-x)^{C_{u_{v}}} \cdot (1 + D_{u_{v}}x + E_{u_{v}}x^{2}) \;\;\;\;\; {\rm and} \;\;\;\;\;
xd_{v}(x) \;\;\; =  \;\;\; A_{d_{v}}x^{B_{d_{v}}} \cdot (1-x)^{C_{d_{v}}} \\
x\bar{U}(x) & = & A_{\bar{U}}x^{B_{\bar{U}}} \cdot (1-x)^{C_{\bar{U}}} \;\;\;\;\; {\rm and} \;\;\;\;\;
x\bar{D}(x) \;\;\; = \;\;\; A_{\bar{D}}x^{B_{\bar{D}}} \cdot (1-x)^{C_{\bar{D}}}
\end{eqnarray*}
In the $14$ parameter fit $A_{g}$, $A_{u_{v}}$ and $A_{d_{v}}$ are constrained by the momentum and quark sum rules. Furthermore it is assumed that $B_{\bar{U}} = B_{\bar{D}}$, $C_{g}^{\prime}=25$ and $A_{\bar{U}} = A_{\bar{D}}(1-f_{s})$, where $f_{s}$ is the strange sea fraction at $Q_{0}^{2}$. These choices define the central fit, as adding more parameters does not significantly reduce the $\chi^{2}$. The parametrization uncertainty is estimated by adding one additional parameter at a time, changing the parametrizations and varying the starting scale $Q_{0}^{2}$ around the default value of $1.9$~GeV$^{2}$. A model uncertainty is determined by varying the heavy quark masses, $f_{s}$ and $Q^{2}_{\rm min}$ around their default values. $Q^{2}_{\rm min}$ specifies the minimal $Q^{2}$ for a data point to be included in the fit.   

The fit of the PDFs is performed using the NLO DGLAP equations, employing the QCDNUM \cite{qcdnum17} evolution code. Charm and beauty quarks are treated as massive in the Thorne-Roberts variable-flavor-number scheme \cite{rt-hfl,rt-vfns}. The jet production cross sections are calculated with the fastNLO code \cite{fastnlo10}, which efficiently convolutes the matrix elements, calculated using the NLOJet++ program \cite{nlojet++}, with the fitted PDFs. 

Here is not the place to discuss this, but it should be mentioned that HERAPDF1.0 and HERAPDF1.5 fits have been also performed using evolution equations in next-to-next-to-leading order (NNLO). For the reader who wants to know more details and updates on the HERAPDFS, they can be found here \cite{herapdftable}.

\section{Data}
The data used in the extraction of the PDFs and the strong coupling $\alpha_{S}(M_{Z})$ are the combined NC and CC $e^{+}p$ and $e^{-}p$ inclusive cross sections from the HERA-I \cite{herapdf10,yoshida} and HERA-II running periods \cite{h1-zeus-data-comb-prel,yoshida}. In addition, the following jet measurements from the H1 and ZEUS collaborations were used: normalized inclusive jet cross sections for $150 < Q^{2} < 15000$~GeV$^{2}$ \cite{h1-jets-highq2-alphas}, corresponding to an integrated luminosity of $395$~pb$^{-1}$, and inclusive jet cross sections for $5 < Q^{2} < 100$~GeV$^{2}$ \cite{h1-jets-lowq2-alphas}, corresponding to an integrated luminosity of $43.5$~pb$^{-1}$, and inclusive jet cross sections from two running periods in HERA-I for $Q^{2} > 125$~GeV$^{2}$ \cite{Chekanov:2002be}, corresponding to an integrated luminosity of $38.6$~pb$^{-1}$, and for $125 < Q^{2} < 10000$~GeV$^{2}$ \cite{Chekanov:2006xr}, corresponding to an integrated luminosity of $82$~pb$^{-1}$.

\section{Fit Results}
\begin{figure}[h]
  \hspace{-5mm}
  \includegraphics[height=.22\textheight]{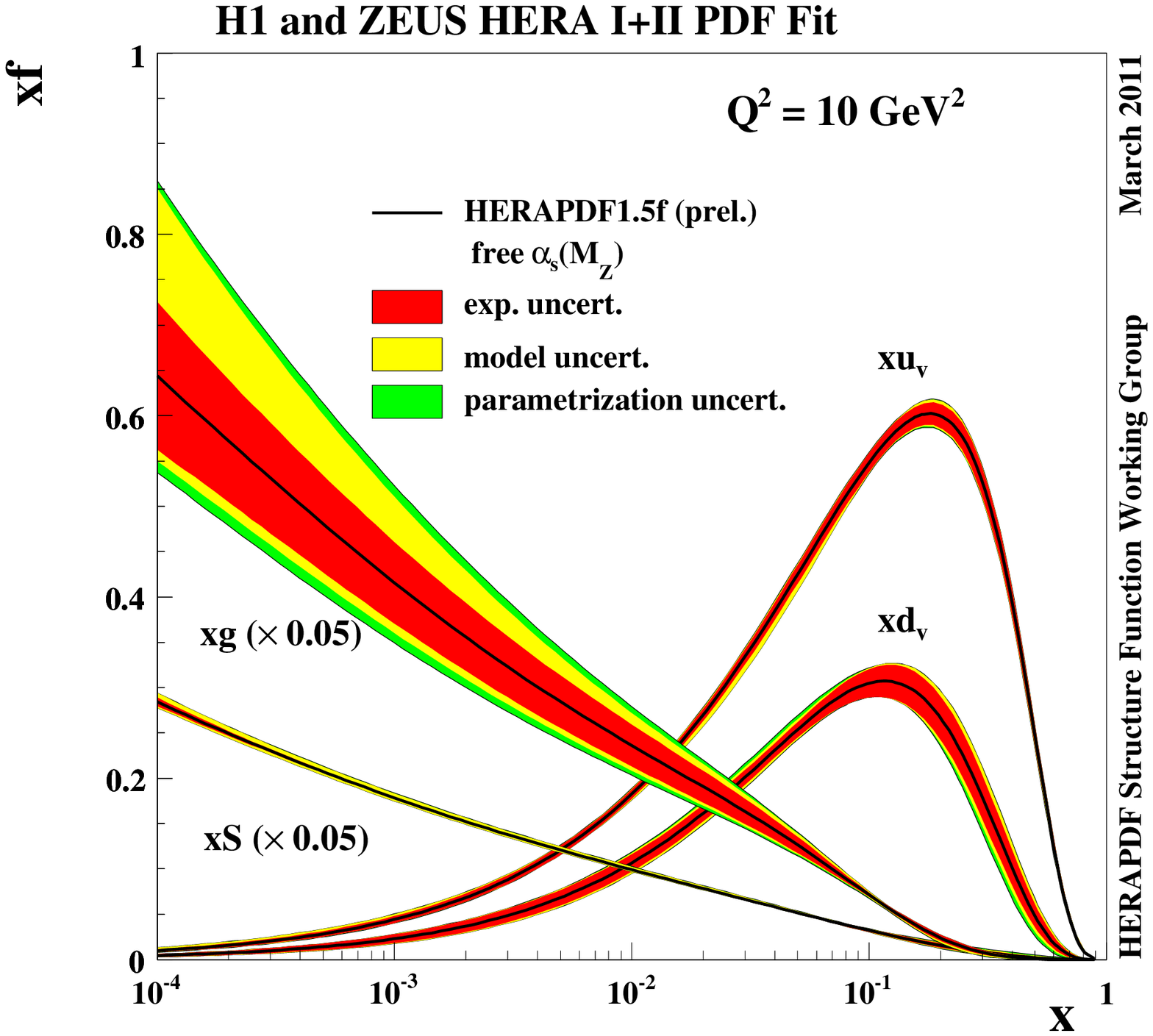}
  \put (-126.0,121.0){\small (a)}
  \hspace{2mm}
  \includegraphics[height=.22\textheight]{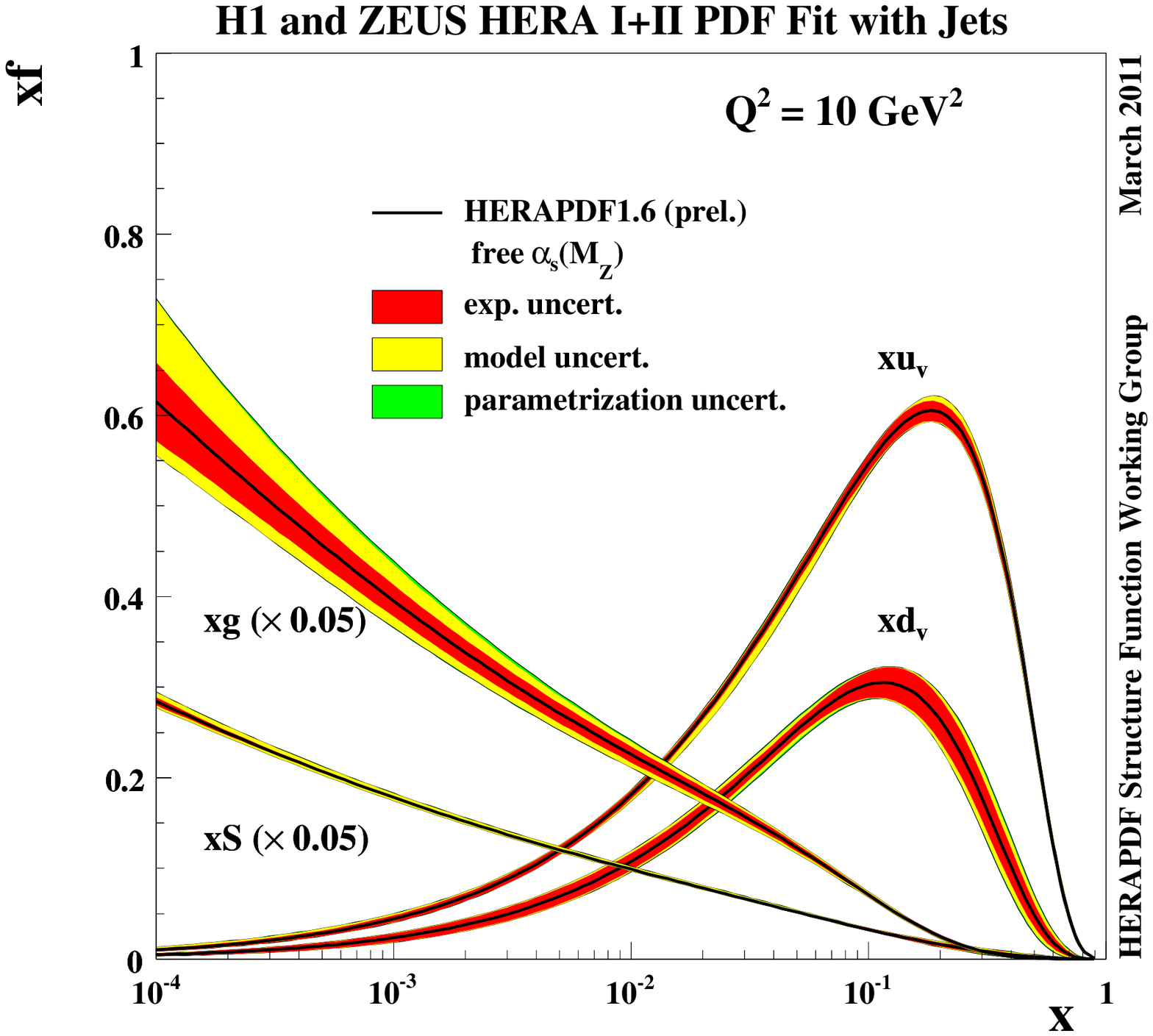}
  \put (-126.0,121.0){\small (b)}
  \hspace{7mm}
  \includegraphics[height=.21\textheight,width=.25\textwidth,bbllx=0pt,bblly=8pt,bburx=560pt,bbury=429pt,clip=]{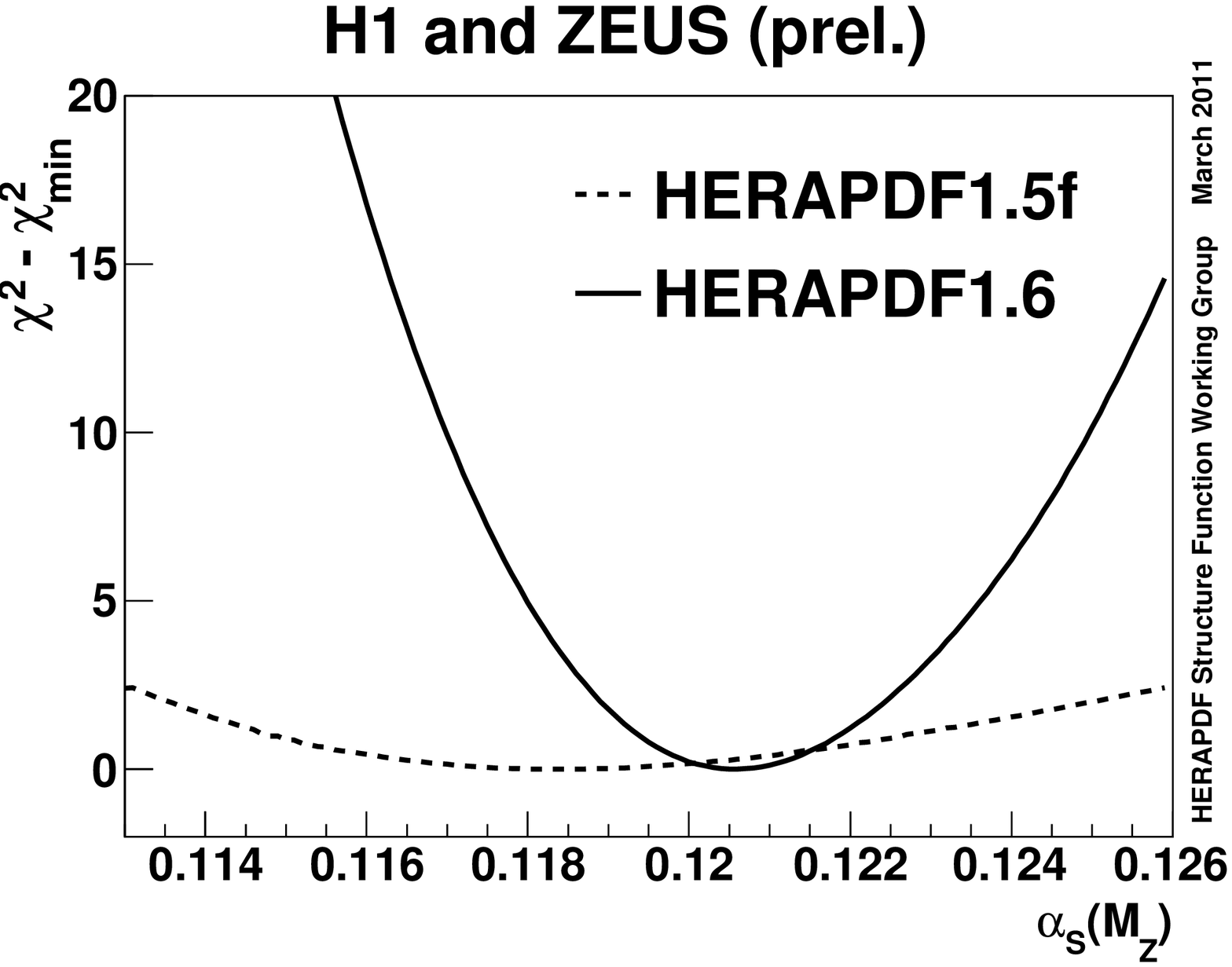}
  \put (-126.0,121.0){\small (c)}
  \caption{
(a) HERAPDF1.5f without and (b) HERAPDF1.6 with jet data included in the fit, as a function of $x$ for $Q^{2}=10$~GeV$^{2}$ with $\alpha_{S}(M_{Z})$ as an additional free parameter in the fit, and (c) $\Delta\chi^{2}$ for both fits.
          }
  \label{fig1} 
\end{figure}

The PDFs as a function of $x$ at $Q^{2}=10$~GeV$^{2}$, when going from HERAPDF1.5 (10 parameters) to HERAPDF1.5f (14 parameters) with fixed $\alpha_{S}(M_{Z})=0.1176$, show an overall similar total uncertainty. The latter yields a softer sea density at high $x$ and a suppressed d-valence density at low $x$. Keeping $\alpha_{S}(M_{Z})$ fixed and including the jet measurements, which leads to the HERAPDF1.6 fit, one observes little difference in the PDFs and their uncertainties, only the high $x$ gluon uncertainty is somewhat reduced. The respective figures can be found in \cite{h1-zeus-pdf-alphas-fits}. More interesting results are obtained when the fits for HERAPDF1.5f (no jets) and HERAPDF1.6 (with jets) are performed with $\alpha_{S}(M_{Z})$ treated as a free fit parameter. The corresponding PDFs and their uncertainty bands are shown in Fig.\ref{fig1}. 
Adding the jet data leads to a dramatic decrease of the low $x$ gluon uncertainty, not only of the experimental uncertainty, but also of the model and parametrization ones.
In Fig.\ref{fig1}c the $\Delta\chi^{2}$ distributions are shown as a function of $\alpha_{S}(M_{Z})$. The HERAPDF1.5f fit shows a shallow minimum, while the HERAPDF1.6 fit provides a strong constraint on the value of $\alpha_{S}(M_{Z})$. The addition of the jet data successfully reduces the correlation of the strong coupling and the gluon density, for reasons discussed in the introduction. The increase in the uncertainty of the gluon density due to $\alpha_{S}(M_{Z})$ being a free parameter is not large when jet data are included. The fit yields the following value for the strong coupling:
\begin{floatingfigure}[r]{68mm}
  \vspace{6mm}
  \mbox{ 
  \includegraphics[width=.35\textwidth,height=.25\textheight,bbllx=20pt,bblly=10pt,bburx=562pt,bbury=478pt,clip=]{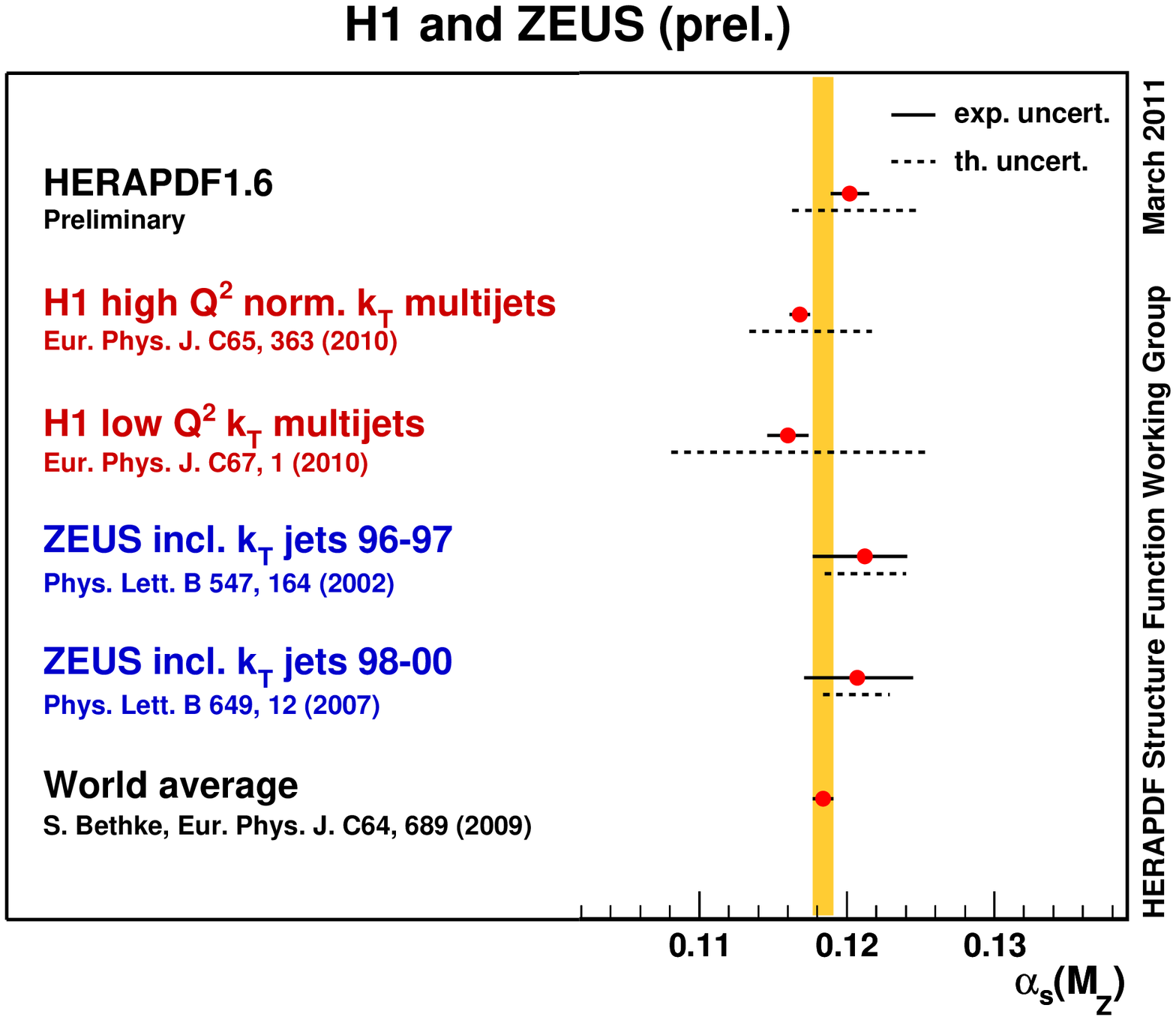}
       } 
  \centerline{ {\bf\small FIGURE 2.} \hspace{0mm} $\alpha_{S}(M_{Z})$ values from HERAPDF1.6, } 
  \centerline{ \hspace{-9mm} from jet production in DIS, by the H1 and } 
  \centerline{ \hspace{-5mm} ZEUS collaborations, and the world average. }   
  \label{fig3} 
\end{floatingfigure}
\quad
\begin{eqnarray*}
\hspace*{0mm}  \alpha_{S}(M_{Z}) & = & 0.1202 \pm 0.0013 \, {\rm (exp.)} \pm 0.0007 \, {\rm (model/param.)}  \\
&  & \pm 0.0012 \, {\rm (hadronization)} \, ^{+0.0045}_{-0.0036} \, {\rm (scale)} 
\end{eqnarray*}
The hadronization uncertainty is due to the corresponding corrections of the NLO jet cross sections. The uncertainty
due to the scale is estimated by a variation of the renormalization and factorization scale by the conventional factor of
$1/2$ and $2$ when calculating the jet cross sections. For the inclusive cross sections this uncertainty was assumed to be negligible. 

In Fig.2 the results on $\alpha_{S}(M_{Z})$ are displayed from HERAPDF1.6 and as obtained from recent extractions by H1 and ZEUS using only jet production data in DIS. The experimental and theoretical uncertainties are indicated separately. For HERAPDF1.6 the uncertainty due to the PDFs is part of the experimental uncertainty, while for the jets only results it is part of the theory uncertainty. The values obtained using different methods and experiments are experimentally very precise. They are consistent with each other and with the world average.

\section{Summary}
A NLO QCD fit with simultaneous determinations of the proton PDFs and $\alpha_{S}(M_{Z})$ was performed using HERA DIS data only. Combined H1 and ZEUS inclusive NC and CC cross sections together with inclusive jet cross sections from the H1 and ZEUS collaborations have been used. Including jet data in the fit and treating $\alpha_{S}(M_{Z})$ as a free parameter, dramatically reduces the correlations between the gluon density and the strong coupling compared to the fit without jets. The precision of the gluon density is improved, and an accurate and unbiased determination of $\alpha_{S}(M_{Z})$ is achieved, yielding a value consistent with the world average.





\begin{thebibliography}{99}

\bibitem{h1-contact-interactions}
  F.~D.~Aaron {\it et al.}  [H1 Collaboration],
  accepted by \emph{Phys.\ Lett.\ } \textbf{B}, (2011).
  [arXiv:1107.2478 [hep-ex]].
  
\bibitem{herapdf10}
  F.~D.~Aaron {\it et al.} [H1 and ZEUS Collaboration],
  \emph{JHEP} \textbf{1001}, 109 (2010).
  [arXiv:0911.0884 [hep-ex]].

\bibitem{herapdf15} 
  H1 and ZEUS Collaboration, 
  ``PDF fits including HERA-II high $Q^{2}$ data,'' 
  \emph{H1prelim-10-142}, \emph{ZEUS-prel-10-018} (2010).

\bibitem{h1-zeus-pdf-alphas-fits} 
  H1 and ZEUS Collaboration, 
  ``QCD analysis and determination of $\alpha_{S}(M_{Z})$ using the combined H1 and ZEUS inclusive cross sections together with the jet production cross section measured by the H1 and ZEUS experiments,'' 
  \emph{H1prelim-11-034}, \emph{ZEUS-prel-11-001}.

\bibitem{qcdnum17}
  M.~Botje,
  \emph{Comput.\ Phys.\ Commun.\ } \textbf{182}, 490 (2011).
  [arXiv:1005.1481 [hep-ph]].

\bibitem{rt-hfl}
  R.~S.~Thorne and R.~G.~Roberts,
  \emph{Phys.\ Rev.\ } \textbf{D57}, 6871 (1998).
  [hep-ph/9709442].

\bibitem{rt-vfns}
  R.~S.~Thorne,
  \emph{Phys.\ Rev.\ } \textbf{D73}, 054019 (2006). 
  [hep-ph/0601245].
  
\bibitem{fastnlo10}
  T.~Kluge, K.~Rabbertz, M.~Wobisch,
  ``FastNLO: Fast pQCD calculations for PDF fits,''
  [hep-ph/0609285].  

\bibitem{nlojet++}
  Z.~Nagy, Z.~Trocsanyi,
  \emph{Phys.\ Rev.\ Lett.\ }  \textbf{87}, 082001 (2001).
  [hep-ph/0104315].
  
\bibitem{herapdftable}
  H1 and ZEUS Collaboration,
  {\small\url{https://www.desy.de/h1zeus/combined_results/herapdftable/}}  
  
\bibitem{yoshida}
  R.~Yoshida,  
  ``Combined Measurements of NC and CC DIS Cross-Sections at HERA,''
  these proceedings.
  
\bibitem{h1-zeus-data-comb-prel} 
  H1 and ZEUS Collaboration, 
  ``Combined Measurement of Neutral and Charged Current Cross Sections at HERA,'' 
  \emph{H1prelim-10-141}, \emph{ZEUS-prel-10-017}.
  
\bibitem{h1-jets-highq2-alphas}
  F.~D.~Aaron {\it et al.}  [H1 Collaboration],
  \emph{Eur.\ Phys.\ J.\ } C \textbf{65}, 363 (2010).
  [arXiv:0904.3870 [hep-ex]].

\bibitem{h1-jets-lowq2-alphas}
  F.~D.~Aaron {\it et al.} [H1 Collaboration],
  \emph{Eur.\ Phys.\ J.\ } C \textbf{67}, 1 (2010).
  [arXiv:0911.5678 [hep-ex]].

\bibitem{Chekanov:2002be}
  S.~Chekanov {\it et al.} [ZEUS Collaboration],
  \emph{Phys.\ Lett.\ } \textbf{B547}, 164 (2002).
  [hep-ex/0208037].
  
\bibitem{Chekanov:2006xr}
  S.~Chekanov {\it et al.} [ZEUS Collaboration],
  \emph{Nucl.\ Phys.\ } \textbf{B765}, 1 (2007).
  [hep-ex/0608048].  
  
  
\end{thebibliography}
\end{document}